\documentclass[prd,twocolumn,amsmath,amssymb]{revtex4}

\usepackage{amsmath,amsfonts,amssymb}
\usepackage{graphicx}
\usepackage[english]{babel} 
\usepackage{color}
\usepackage{booktabs}
\usepackage{feynmp}

\usepackage{graphics}
\usepackage{rotate}
\usepackage{fancyhdr} 

\usepackage{appendix}

\def\VEV#1{\left\langle #1 \right\rangle}

\newcommand{\nn}{\nonumber}
\newcommand{\be}{\begin{equation}}
\newcommand{\ee}{\end{equation}}
\newcommand{\bga}{\begin{gathered}}
\newcommand{\ega}{\end{gathered}}
\newcommand{\beqa}{\begin{eqnarray}}
\newcommand{\eeqa}{\end{eqnarray}}

\newcommand{\bfx}{\mathbf{x}}
\newcommand{\bfn}{\mathbf{\hat{n}}}
\newcommand{\bfgk}{\mathbf{K}}
\newcommand{\bfk}{\mathbf{k}}
\newcommand{\lmax}{l_{\mathrm{max}}}

\def\CG#1#2#3#4#5#6{ \VEV{#1#2#3#4|#5#6}}

\def\wigner#1#2#3#4#5#6{ \mathcal{W}^{#1#3#5}_{#2#4#6}}

\begin{document}

\title{Seeking Inflation Fossils in the Cosmic Microwave Background}
\author{Liang Dai, Donghui Jeong, and Marc Kamionkowski}
\affiliation{Department of Physics \& Astronomy, Bloomberg Center, The Johns Hopkins University, Baltimore, MD 21218, USA}

\date{\today}

\begin{abstract}

If during inflation the inflaton couples to a ``fossil'' field, some new scalar, vector, or tensor field, it typically induces a scalar-scalar-fossil bispectrum. Even if the fossil field leaves no direct physical trace after inflation, it gives rise to correlations between different Fourier modes of the curvature or, equivalently, a nonzero curvature trispectrum, but without a curvature bispectrum. Here we quantify the effects of a fossil field on the cosmic microwave background (CMB) temperature fluctuations in terms of bipolar spherical harmonics (BiPoSHs). The effects of vector and tensor fossils can be distinguished geometrically from those of scalars through the parity of the BiPoSHs they induce. However, the two-dimensional nature of the CMB sky does not allow vectors to be distinguished geometrically from tensors. We estimate the detectability of a signal in terms of the scalar-scalar-fossil coupling for scalar, vector, and tensor fossils, assuming a local-type coupling. We comment on a divergence that arises in the quadrupolar BiPoSH from the scalar-scalar-tensor correlation in single-field slow-roll inflation.

\end{abstract}
\maketitle

\section{Introduction}
\label{sec:intro}

Despite the observational success of the inflation paradigm, the
nature of the inflationary epoch is still poorly understood. The
single-field slow-roll model, being the simplest inflation
model, might not be the whole story of the physics behind
inflation. An abundance of inflation models introduce new
fields, which are often coupled to the
inflaton~\cite{InflationModelReview,DBI-Inflation}. Candidates
for those new degrees of freedom include extra scalar fields in
extensions of the simplest inflation
model~\cite{InflationExtraScalar}. They might also be primordial
vector fields that drive or impact
inflation~\cite{InflationExtraVector}, or even new vectorial
degrees of freedom arising from modified theories of
gravity~\cite{Jimenez:2009ai}. Even in the simplest single-field
slow-roll inflation, tensor metric perturbations \cite{InflationGW} couple to the inflaton~\cite{Maldacena:2002vr,Seery:2008ax,Giddings:2010nc,Giddings:2011zd}.

Fortunately, the cosmic microwave background (CMB) provides a special window to probe physics at very early moments, by encoding non-trivial primordial correlations beyond Gaussianity in its anisotropy pattern. By measuring the primordial bispectra, trispectra, and possibly higher-order correlation functions from its temperature and polarization anisotropies~\cite{Komatsu:2001rj,Babich:2004yc,Bartolo:2004if,Kogo:2006kh,Smidt:2010ra}, various inflation models will hopefully be distinguished, and new physics might be revealed.

Recently, Ref.~\cite{Jeong:2012df} proposed a generic parameterization of the primordial bispectra involving the scalar metric perturbation $\Phi$ (the gauge-invariant Bardeen potential~\cite{Bardeen:1980kt}) and a new field $h$ from inflation. The new field is dubbed an {\it inflation fossil}, a hypothesized primordial degree of freedom that no longer interacts or very weakly interacts during late-time cosmic evolution. The only observational effect of an inflation fossil might therefore be its imprint in the primordial curvature perturbation. Inflation fossils can be those extra fields that are introduced in a variety of alternatives to the single-field slow-roll model. Different models may be distinguished by the spin of the new field. To treat different possibilities for the spin in a model-independent fashion, in Ref.~\cite{Jeong:2012df} the fossil field is parameterized by a symmetric traceless tensor field. Since a symmetric traceless tensor contains a longitudinal scalar (L), two divergence-free vectors (V), and two divergence-free traceless tensors (T), the parameterization allows for all three possibilities for the spin.

\begin{figure}[ht]
\centering
\includegraphics[scale=0.6]{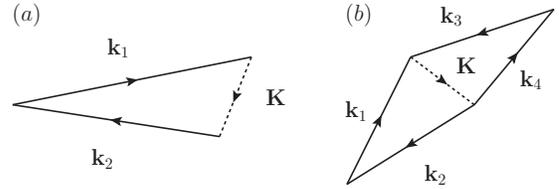}
\caption{Primordial correlations of scalar perturbation (solid vectors) induced by the fossil field (dashed vectors) in Fourier space: (a) For a given realization of the fossil field, two different scalar-perturbation modes are correlated with each other. (b) For a stochastic background of fossil fields, a connected trispectrum of scalar perturbations is induced.}
\label{fig:4pt-correlation}
\end{figure}

A given realization of the fossil field causes two different scalar perturbation modes to correlate with each other (panel (a) of Fig.~\ref{fig:4pt-correlation}). Scalar, vector, and tensor fossils give rise to geometrically distinct cross-correlations. For a stochastic fossil background, connected trispectra that correlate four differents modes of $\Phi$ are generated (panel (b) of Fig.~\ref{fig:4pt-correlation}) when an ensemble average over all fossil-field realizations is performed. Such a scenario of scalar four-point correlations but without scalar three-point correlations can be sought in galaxy surveys where the correlations from scalar, vector and tensor fossils can be disentangled geometrically~\cite{Jeong:2012df}.

In this paper we study the CMB signatures of fossil
fields. Correlations between different Fourier modes of the
curvature perturbation induce couplings between different
CMB-temperature spherical-harmonic coefficients. We parametrize
these cross-correlations in terms of bipolar spherical harmonics
(BiPoSHs). In the presence of some realization of the fossil
field, the CMB appears non-Gaussian with nonvanishing BiPoSHs,
or equivalently, a nonzero four-point correlation function, or
trispectrum. Here we first calculate the BiPoSH coefficients
from a specific realization of the fossil field, and then find
the BiPoSH power spectra for a stochastic background for the
fossil field. We write down the minimum-variance CMB estimators
for the amplitudes of the fossil-field power spectra. As an
example, we numerically examine the case of a local-type
scalar-scalar-fossil bispectrum with a scale-free fossil
spectrum. That case can be described by two parameters, the
amplitude of the bispectrum, and the normalization of the fossil
power spectrum. We find that the dominant effects from a
local-type bispectrum are modulations of small-angular-scale
correlations on large angular scales. We evaluate the
sensitivities for the reduced amplitude of the fossil
background, a combination of the two parameters, for different
fossil-field spin. Our results show that the sensitivity
achievable by Planck is more than an order of magnitude better
than that by current galaxy-clustering surveys and is comparable
to that of larger next-generation surveys.

This paper is organized as the following. In Sec.~\ref{sec:cmb-biposh-inflation-fossils} we first review the parametrization of the effect of fossil fields on scalar perturbations. Since this paper is concerned with the calculation of observables on a spherical sky, we recast the parametrization in Ref.~\cite{Jeong:2012df} in terms of total-angular-momentum (TAM) waves~\cite{Dai:2012bc}. The TAM formalism makes the rotational symmetry of the observed sky manifest throughout and thus greatly simplifies our calculation. We then move on to present the calculation of the BiPoSHs and the BiPoSH power spectra. After that, we construct the minimal-variance quadratic estimator for the reduced amplitude, and calculate the cumulative signal-to-noise. We proceed to present numerical results for a local scalar-scalar-fossil bispectrum in Sec.~\ref{sec:numerical}. They are complemented by Sec.~\ref{sec:local-modulation-real-space}, which discusses a real-space picture of the fossil-field effects on the temperature map. Finally, we make concluding remarks in Sec.~\ref{sec:conclusion}. We also comment there on a divergence in the predicted observables from the scalar-scalar-tensor correlation in single-field slow-roll inflation.

\section{CMB BiPoSHs from inflation fossils}
\label{sec:cmb-biposh-inflation-fossils}

\subsection{Fossil parameterization}
\label{subsec:fossil-para}

Interactions between a fossil field $h$ and the curvature perturbation $\mathcal{R}$ can generate three-point functions of the type $\langle\mathcal{R}\mathcal{R} h\rangle$ at horizon crossing that then convert into a primordial bispectrum involving two scalar-metric-perturbation $\Phi$ modes and a fossil-field mode after the end of inflation. A realization of the $h$ field then locally induces departures from statistical homogeneity in the scalar autocorrelation function, which then appears as a correlation between different Fourier modes of the scalar perturbation. A general parameterization,
\begin{align}
\label{eq:JK-fourier}
\langle \Phi(\bfk_1) \Phi(\bfk_2) \rangle_{h^{p}(\bfgk)} = & (2\pi)^3 \delta_D (\bfk_1+\bfk_2+\bfgk) f^{p}_h(k_1,k_2,K) \nn\\
                                                           & \times \epsilon^{p}_{ab}(\bfgk) \hat{k}^a_1 \hat{k}^b_2 [h^{p}(\bfgk)]^*.
\end{align}
is written down in Ref.~\cite{Jeong:2012df} to account for the possibility that the fossil field $h$ can be scalar, vector or tensor field. Here $\bfk_1$ and $\bfk_2$ are wavevectors of the two scalar-perturbation modes, $\bfgk$ is the wavevector of the fossil-field mode, and $h^{p}(\bfgk)$ is the Fourier amplitude of the fossil field. The Dirac delta function reflects the translational invariance of the underlying physics. We introduce a generic bispectrum shape function $f^{p}_h(k_1,k_2,K)$, which is related to the primordial scalar-scalar-fossil bispectrum through $B_{\Phi\Phi h}(\bfk_1,\bfk_2,\bfgk)=P^p_h(K)f^p_h(k_1,k_2,K)\epsilon^{p}_{ab}(\bfgk) \hat{k}^a_1 \hat{k}^b_2$, where $P^p_h(K)$ is the power spectrum for the fossil field. Moreover, the symmetric three-by-three polarization tensor $\epsilon^{p}_{ab}(\bfgk)$ with $p=0,L,x,y,+,\times$ geometrically distinguishes modulations of scalar, vector, and tensor type through the azimuthal dependence when the configuration is rotated about the direction of $\bfgk$. The trace tensor $\epsilon^0_{ab}(\bfgk) \propto \delta_{ab}$ and the longitudinal $\epsilon^{L}_{ab}(\bfgk)\propto (K_a K_b - K^2 \delta_{ab}/3)K^{-2}$ describe a scalar fossil field. Two transverse-vectorial tensors $\epsilon^{x,y}_{ab}(\bfgk) \propto K_{(a}w^{x,y}_{b)}$ satisfying $K^a w^{x,y}_a=0$ describe a transverse-vector fossil field. Likewise, two transverse-tensorial tensors $\epsilon^{+,\times}_{ab}(\bfgk)$ satisfying $K^a\epsilon^{+,\times}_{ab}(\bfgk)=0$ describe a transverse-tensor fossil field.

\subsection{The TAM formalism}
\label{subsec:tam-formalism}

Since angular observables defined on the two-dimensional sky are involved, we adopt the total-angular-momentum (TAM) wave formalism, recently developed in Ref.~\cite{Dai:2012bc}, to take full advantage of the rotational symmetry of the problem from the very beginning. In the TAM formalism, scalar, vector, or tensor fields in three-dimensional space are expanded in terms of a complete set of spherical waves, as opposed to the conventional expansion in terms of plane waves. These spherical waves are eigenfunctions of the Helmholtz equation with wave number $K$ and are eigenfunctions of total angular momentum and its third component with quantum numbers $J$ and $M$. Although the scalar case has been long known as the Fourier-Bessel expansion, the TAM formalism greatly simplifies calculations when vector or tensor fields have to be projected onto the two-dimensional sphere. In our case here, since three-by-three symmetric polarization tensors are used in Eq.~(\ref{eq:JK-fourier}), scalar, vector and tensor fossil fields can all be incorporated into a symmetric traceless tensor field $h_{ab}(\bfx)$. As the tensor field is expanded in terms of TAM waves, the longitudinal mode $h^L_{JM}(K)$ describes a scalar fossil field, the two divergence-free vectorial modes $h^{VE}_{JM}(K)$ and $h^{VB}_{JM}(K)$ describe a transverse-vector fossil, and the two divergence-free tensorial modes $h^{TE}_{JM}(K)$ and $h^{TB}_{JM}(K)$ describe a transverse-tensorial fossil. 

In terms of TAM coefficients $\Phi_{lm}(k)$ and $h^{\alpha}_{JM}(K)$ for the scalar perturbation and the fossil field respectively, the modulation analogous to Eq.~(\ref{eq:JK-fourier}) reads
\begin{align}
\label{eq:JK-tam}
&\langle \Phi_{l_1 m_1}(k_1) \Phi_{l_2 m_2}(k_2) \rangle_{h^{\alpha}_{JM}(K)} = [h^{\alpha}_{JM}(K)]^* f^{\alpha}_h(k_1,k_2,K)\nn\\
&\times  (4\pi)^3 (-i)^{l_1+l_2+J}\frac{1}{k_1 k_2}\nn\\
&\times \int d^3\bfx \left(\nabla^a \Psi^{k_1}_{(l_1 m_1)}(\bfx)\right) \left(\nabla^b \Psi^{k_2}_{(l_2 m_2)}(\bfx)\right) \Psi^{\alpha,K}_{(JM)ab}(\bfx),
\end{align}
for $\alpha=L,VE,VB,TE,TB$ respectively. Here $\Psi^{k}_{(lm)}(\bfx)$ and $\Psi^{\alpha,K}_{(JM)ab}(\bfx)$ are TAM wave functions for scalar and tensor fields. The overlaps of three TAM wave functions have been worked out in Ref.~\cite{Dai:2012ma}. Note that statistical homogeneity and isotropy require that $f^{VE}_h(k_1,k_2,K)=f^{VB}_h(k_1,k_2,K) \equiv f^{V}_h(k_1,k_2,K)$ and $f^{TE}_h(k_1,k_2,K)=f^{TB}_h(k_1,k_2,K) \equiv f^{T}_h(k_1,k_2,K)$. 

\subsection{CMB BiPoSHs}
\label{subsec:cmb-biposh}

Through the epochs of radiation domination and matter domination to recombination, scalar metric perturbations source temperature and $E$-mode polarization anisotropies of the CMB. Therefore, cross-correlations of different scalar-perturbation modes, as a consequence of modulation by the fossil field, give rise to cross-correlations between different harmonic modes of the CMB anisotropies. For clarity, in this paper we only discuss the effect on the temperature map. Still, the inclusion of $E$-mode polarization is straightforward and will improve the overall sensitivity of detection, once the cross-correlations between temperature and $E$-mode polarization are properly taken care of.

For a given fossil-field configuration, the modulation violates statistical isotropy in the anisotropies. This effect can be conveniently quantified in terms of a bipolar spherical harmonic (BiPoSH) expansion~\cite{Hajian:2004zn}. Being the two-sphere analog of the fossil-field modulation, Eq.~(\ref{eq:JK-fourier}), it provides a parameterization for the most general two-point correlation function on the sky. Specifically, cross-correlations of CMB temperature multipoles read
\begin{align}
\left\langle a^T_{l_1 m_1} a^{T*}_{l_2 m_2} \right\rangle_h & =  C^{TT}_{l_1}\delta_{l_1 l_2}\delta_{m_1 m_2} \nn\\
& + \sum_{JM} (-1)^{m_2} \langle l_1 m_1 l_2,-m_2 |J M \rangle A^{JM}_{l_1 l_2},
\end{align}
where $\langle l_1 m_1 l_2 m_2|l_3 m_3\rangle$ denotes the Clebsch-Gordan coefficient. The BiPoSH coefficients $A^{JM}_{l_1 l_2}$, given by
\begin{align}
\label{eq:biposh-A}
A^{JM}_{l_1 l_2} & =  (-1)^{l_1+l_2+M} \sqrt{2J+1}\nn\\
&\times \sum_{m_1 m_2} (-1)^{m_2} \wigner{l_1}{m_1}{l_2}{,-m_2}{J}{,-M} \left\langle a^T_{l_1 m_1} a^{T*}_{l_2 m_2} \right\rangle_h,
\end{align}
exist if $J$, $l_1$ and $l_2$ can form a triangle. The angle brackets with a subscript $h$ denote an average over all realizations of $\Phi$ for a fixed realization of $h$. Here we use $\wigner{l_1}{m_1}{l_2}{m_2}{l_3}{m_3}$ as a compact notation for the usual Wigner-3$j$ symbol. Since the fossil is parameterized by a symmetric tensor field, it only induces BiPoSHs with $J\geqslant 2$.

Under the TAM basis, the scalar perturbation can be described by TAM coefficients $\Phi_{lm}(k)$, which source CMB temperature multipoles according to
\be
\label{eq:temp-multipole}
a^{T}_{lm} = \frac{1}{2\pi^2} (-i)^l \int k^2 dk g^T_l(k)\Phi_{lm}(k),
\ee
where $g^T_l(k)$ is the scalar radiation transfer function for temperature.

Combining Eqs.~(\ref{eq:JK-tam}), (\ref{eq:biposh-A}), and (\ref{eq:temp-multipole}), the CMB BiPoSH due to modulation of a single TAM wave of the fossil field is,
\begin{align}
\label{eq:biposh-A-result}
&\left. A^{JM}_{l_1 l_2} \right|_{h^{\alpha}_{JM}(K)} = -(-i)^{J} (-1)^{l_1+l_2+P(\alpha)} h^{\alpha}_{JM}(K) \frac{16}{\pi} \nn\\
&\times \left(\frac{(2l_1+1)(2l_2+1)}{4\pi}\right)^{\frac{1}{2}} \int k_1^2 dk_1 g^T_{l_1}(k_1) \int k_2^2 dk_2 g^T_{l_2}(k_2)\nn\\
&\times  f^{\alpha}_h(k_1,k_2,K) \mathcal{I}^{\alpha}_{l_1 l_2 J}(k_1,k_2,K), 
\end{align}
where the parity $P(\alpha)=0$ for $\alpha=L,VE,TE$ and $P(\alpha)=1$ for $\alpha=VB,TB$. The functions $\mathcal{I}^{\alpha}_{l_1 l_2 J}(k_1,k_2,K)$ are given by
\begin{align}
\label{eq:radial-integral-def}
& \mathcal{I}^{\alpha}_{l_1 l_2 J}(k_1,k_2,K) = \left[\frac{4\pi}{(2l_1+1)(2l_2+1)(2J+1)}\right]^{1/2} \nn\\
&\times \left[ \int d^3\bfx \Psi^{L,k_1,a}_{(l_1 m_1)}(\bfx) \Psi^{L,k_2,a}_{(l_2 m_2)}(\bfx) \Psi^{\alpha,K}_{(JM)ab}(\bfx) \right] / \wigner{J}{M}{l_2}{m_2}{l_1}{m_1},
\end{align}
where the relevant overlaps of three TAM wave-functions can be found in Eqs.~(80), (81), (84), (85), and (87) of Ref.~\cite{Dai:2012ma}, for each $\alpha$ respectively. Note that the dependence on azimuthal quantum numbers $m_1$, $m_2$, and $M$ cancels out on the right hand side, and the final result for $\mathcal{I}^{\alpha}_{l_1 l_2 J}(k_1,k_2,K)$ is reduced to an integral over the radial coordinate.

In the above result for the BiPoSH coefficients, rotational invariance is manifest in that a given $A^{JM}_{l_1 l_2}$ is only generated by TAM waves of the fossil field with the same total-angular-momentum quantum numbers $J$ and $M$. Moreover, due to parity conservation, $L$, $VE$, and $TE$ modes only induce even-parity BiPoSHs, i.e. $J+l_1+l_2=$ even, while $VB$ and $TB$ modes only induce odd-parity BiPoSHs, i.e. $J+l_1+l_2=$ odd. Therefore, vector and tensor fossils, both containing $B$-mode TAM waves, can be distinguished from scalar fossils from their signature in odd-parity BiPoSHs. Nevertheless, vector and tensor fossils cannot be geometrically distinguished from each other from CMB BiPoSHs, as they can with three-dimensional surveys~\cite{Jeong:2012df}. This is heuristically understood as information loss when the three-dimensional correlation function of the scalar perturbation is projected onto the sky to give two-dimensional angular correlation functions of the CMB.

In addition to primordial mechanisms, late-time effects can distort a Gaussian, statistically isotropic map as well. Particularly, weak-lensing of the CMB by the foreground matter distribution also produces even-parity BiPoSHs, mimicking the effect of inflation fossils. Still, we anticipate the spectra of BiPoSHs from lensing to differ from those from modulation by fossils. The reason is that the power spectrum for lensing by the scalar potential peaks at $J\sim 60$~\cite{Book:2011na}, but, e.g., for a local scalar-scalar-fossil bispectrum the fossil modulation effects dominate at $J\lesssim 5$, as our results will show later. This implies that the shape of BiPoSH power spectra can be used to break the degeneracy. Besides, weak-lensing generates $B$-mode polarization of the CMB, while a primordial scalar-scalar-fossil bispectrum does not. 

\subsection{BiPoSH power spectra and estimators}
\label{subsec:biposh-power}

A stochastic background of the fossil field is presumably generated during inflation, just like the scalar field or inflationary gravitational waves, and it is characterized by a power spectrum,
\begin{align}
\left\langle h^{\alpha}_{JM}(K) h^{\alpha'}_{J'M'}(K') \right\rangle & = P^{\alpha}_h(K) \frac{(2\pi)^3}{K^2} \delta_D(K-K')\nn\\
& \times \delta_{JJ'}\delta_{MM'}\delta_{\alpha\alpha'}.
\end{align}
Statistical homogeneity and isotropy guarantee that $P_h^{VE}(K)=P_h^{VB}(K)\equiv P^V_h(K)$ and $P_h^{TE}(K)=P_h^{TB}(K)\equiv P^T_h(K)$.

The statistical significance with which any individual BiPoSH is detected to be nonzero is expected to be small. In light of this, we average over all realizations of the fossil field, and define bipolar auto-/cross-power spectra,
\be
C^{J}_{l_1 l_2,l_3 l_4} = \frac{1}{2J+1} \left\langle \sum_{M=-J}^{J} A^{JM}_{l_1 l_2} \left[ A^{JM}_{l_3 l_4} \right]^* \right\rangle,
\ee
to statistically measure the imprint of fossils. These correspond to four-point correlations in the temperature map. They are invariant under rotations. We emphasize that since the fossil-field power spectrum is statistically homogeneous and isotropic, statistical isotropy of the CMB is resumed after taking an ensemble average over the fossil field.

To measure BiPoSHs from data, we use quadratic estimators. Estimators for BiPoSH coefficients can be constructed by
\be
\widehat{A^{JM}_{l_1 l_2}} = \sum_{m_1 m_2} (-1)^{m_2} \CG{l_1}{m_1}{l_2}{,m_2}{J}{M} a^T_{l_1 m_1}a^{T*}_{l_2 m_2}.
\ee
Then estimators for bipolar power spectra can be written down,
\begin{align}
& \widehat{C^J_{l_1 l_2, l_3 l_4}} = \frac{1}{2J+1} \sum_{M=-J}^J \widehat{A^{JM}_{l_1 l_2}} \left[ \widehat{A^{JM}_{l_3 l_4}} \right]^* \nn\\
& - C^{TT}_{l_1} C^{TT}_{l_2} \left( \delta_{l_1 l_3} \delta_{l_2 l_4} + \delta_{l_1 l_4} \delta_{l_2 l_3} (-)^{l_1+l_2+J} \right).
\end{align}
Note that the second term is needed to unbias the estimators with respect to a Gaussian CMB map without fossil effects.

We assume a phenomenological parameterization for the fossil-field background, which is described by two parameters. One is the normalization $\mathcal{P}^Z_h$ for the power spectrum,
\be
\label{eq:fossil-power}
P^Z_h(K)=\mathcal{P}^Z_h \tilde{P}^Z_h(K),
\ee
with a fiducial-power spectrum shape $\tilde{P}^Z_h(K)$. The other is the amplitude $\mathcal{B}^Z_h$ of the scalar-scalar-fossil bispectrum,
\be
\label{eq:bispectrum-shape}
f^Z_h (k_1, k_2, K) = \mathcal{B}^Z_h \tilde{f}^Z_h (k_1, k_2, K),
\ee
with a fiducial bispectrum shape $\tilde{f}^Z_h (k_1, k_2, K)$. Here $Z$ can be scalar $Z=L$, transverse vector $Z=\{VE,VB\}$ or transverse tensor $Z=\{TE,TB\}$. For vector or tensor fossils, both $E$ modes and $B$ modes have to exist.

With these parameterization, Eq.~(\ref{eq:biposh-A-result}) can be cast into the form
\be
\label{eq:biposh-coef-A}
\left.A^{JM}_{l_1 l_2}\right|_{h^{\alpha}_{JM}(K)} = \mathcal{B}^Z_h F^{J,\alpha}_{l_1 l_2}(K) h^{\alpha}_{JM}(K), 
\ee
where $\alpha\in Z$ and the coefficient function $F^{J,\alpha}_{l_1 l_2}(K)$ can be read off from Eq.~(\ref{eq:biposh-A-result}) as
\begin{align}
\label{eq:Fcoef}
& F^{J,\alpha}_{l_1 l_2}(K) = -(-i)^J (-1)^{l_1+l_2+P(\alpha)} \frac{16}{\pi} \left(\frac{(2l_1+1)(2l_2+1)}{4\pi}\right)^{\frac{1}{2}} \nn\\
& \times \int k_1^2 dk_1 g^T_{l_1}(k_1) \int k_2^2 dk_2 g^T_{l_2}(k_2) \tilde{f}^{Z}_h(k_1, k_2, K) \nn\\
& \times \mathcal{I}^{\alpha}_{l_1 l_2 J}(k_1,k_2,K).
\end{align}
The BiPoSHs power spectra are then calculated to be
\begin{align}
C^{J}_{l_1 l_2,l_3 l_4} & = \frac{\mathcal{A}^{Z}_h}{(2\pi)^3} \sum_{\alpha\in Z} \int K^2 dK \tilde{P}^Z_h(K) F^{J,\alpha}_{l_1 l_2}(K) \left[ F^{J,\alpha}_{l_3 l_4}(K) \right]^* \nn\\
                        & \equiv \mathcal{A}^{Z}_h \mathcal{F}^{J,Z}_{l_1 l_2, l_3 l_4},
\end{align}
where we define the reduced amplitude $\mathcal{A}^Z_h\equiv \mathcal{P}^Z_h(\mathcal{B}^Z_h)^2$ of the fossil background. There is an observational degeneracy between the fossil-field power spectrum and scalar-scalar-fossil bispectrum, so only this combination is measurable. An estimator,
\be
\widehat{\mathcal{A}^{J,Z}_{h,l_1 l_2,l_3 l_4}} = \widehat{C^J_{l_1 l_2,l_3 l_4}} / \mathcal{F}^{J,Z}_{l_1 l_2, l_3 l_4},
\ee
for the reduced amplitude $\mathcal{A}^Z_h$ can be constructed, for each possible combination of $J$ and $l_i,~i=1,\ldots,4$. Taking into account the symmetry of $\widehat{\mathcal{A}^{J,Z}_{h,l_1 l_2,l_3 l_4}}$, without loss of generality we can fix $l_1\leqslant l_2$, $l_3\leqslant l_4$, and $l_2\leqslant l_4$ if $l_1=l_3$.

We can combine these estimators to bring down the statistical error. Under the null hypothesis, it can be shown that $\widehat{\mathcal{A}^{J,Z}_{h,l_1 l_2,l_3 l_4}}$'s, being quartic expressions in $a^T_{lm}$, are nearly uncorrelated with each other. This is seen through calculating the null-hypothesis correlation matrix by applying the Wick expansion. Cross-correlation elements are suppressed by either negative powers of $(2l_i+1)$ or a Wigner-6$j$ symbol relative to the auto-correlation elements. Treating $\widehat{\mathcal{A}^{J,Z}_{h,l_1 l_2,l_3 l_4}}$'s as statistically independent estimators, we then linearly combine them, and write down the inverse-variance-weighted estimator,
\begin{align}
& \widehat{\mathcal{A}^Z_h} = \left\{ \sum_J \sum_{(l_1,l_2,l_3,l_4)} \widehat{\mathcal{A}^{J,Z}_{h,l_1 l_2,l_3 l_4}} \left\langle\left[ \widehat{\mathcal{A}^{J,Z}_{h,l_1 l_2,l_3 l_4}} \right]^2\right\rangle^{-1}_0 \right\} \nn\\
& / \left\{ \sum_L \sum_{(l_1,l_2,l_3,l_4)} \left\langle\left[ \widehat{\mathcal{A}^{J,Z}_{h,l_1 l_2,l_3 l_4}} \right]^2\right\rangle^{-1}_0 \right\},
\end{align}
where $\sum_{(l_1,l_2,l_3,l_4)}$ loops over all {\it independent} combinations of multipoles. The denominator is equal to the inverse variance of $\widehat{\mathcal{A}^Z_h}$, which is worked out to be
\begin{align}
 (\sigma^Z_{\mathcal{A}})^{-2} \equiv \left\langle \left[ \widehat{\mathcal{A}^Z_h} \right]^2 \right\rangle^{-1}_0 = & \frac{1}{8} \sum_J \sum_{l_1 l_2 l_3 l_4} \frac{2J+1}{C^{TT}_{l_1} C^{TT}_{l_2} C^{TT}_{l_3} C^{TT}_{l_4}} \nn\\
& \times \mathcal{F}^{J,Z}_{l_1 l_2,l_3 l_4} \left[ \mathcal{F}^{J,Z}_{l_1 l_2,l_3 l_4} \right]^*.
\end{align}
Hence, a high signal-to-noise $\mathcal{S}=\mathcal{A}^Z_h/\sigma^Z_{\mathcal{A}}$ indicates a detectable BiPoSH signature in the CMB from inflation fossils.

\section{Numerical Results}
\label{sec:numerical}

Our discussion so far applies to the most general inflation fossil with arbitrary power spectrum $P^{\alpha}_h(K)$ and bispectrum shape $f^{\alpha}_h(k_1,k_2,K)$. In order to provide some illustrative numerical estimates of the detectability of the signal, we specialize to the case where the bispectrum shape is of the local type. Such a bispectrum may arise if the fossil-curvature interaction is local. This occurs, for example, for the graviton-curvature-curvature bispectrum~\cite{Maldacena:2002vr} in single-field slow-roll inflation. We approximate the bispectrum in this case by the fiducial form $\tilde{f}_h^Z(k_1,k_2,K)= (k_1 k_2)^{-3/2}$ it attains in the squeezed limit $k_1,k_2 \ll K$ where it peaks. We assume a scale-free power spectrum $\tilde{P}^Z_h(K)=1/K^3$ for the fossil field as should arise if no physical scale other than the Hubble scale comes into play during inflation.

Specializing to those fiducial forms for the power spectrum and the bispectrum, we numerically compute sensivities $\sigma^Z_{\mathcal{A}}$. We assume the standard flat $\Lambda$CDM cosmology with the WMAP+BAO+$H_0$ best-fit cosmological parameters taken from TABLE 1 of Ref.~\cite{Komatsu:2010fb}, except that for consistency we assume perfect scale-invariance during inflation by setting $n_s=1$. The public code $\mathtt{CAMB}$~\cite{Lewis:1999bs} is used to tabulate radiation transfer functions. CMB multipoles up to $\lmax=3000$ are considered.

\begin{figure*}[htbp]
\centering
\hspace{-1cm}
\includegraphics[scale=0.9]{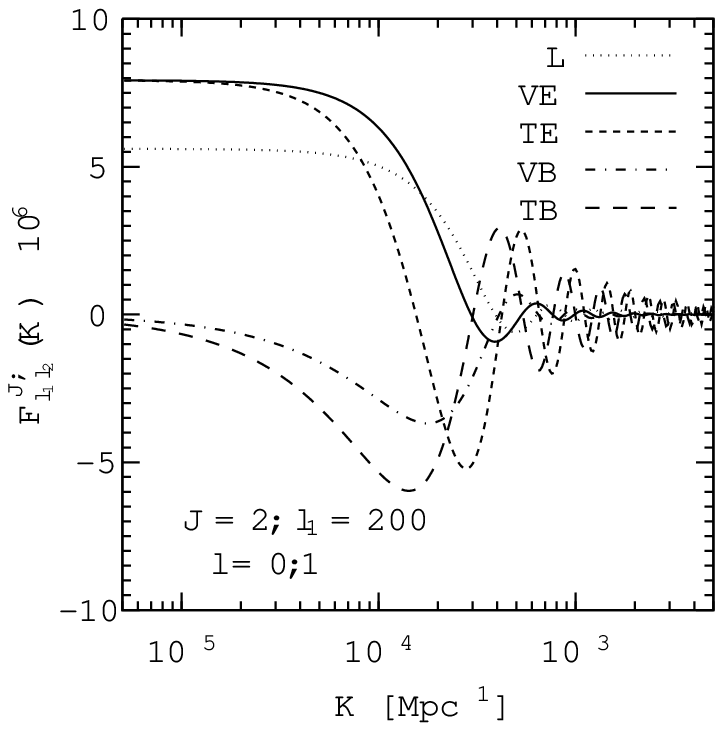}
\hspace{+1cm}
\includegraphics[scale=0.9]{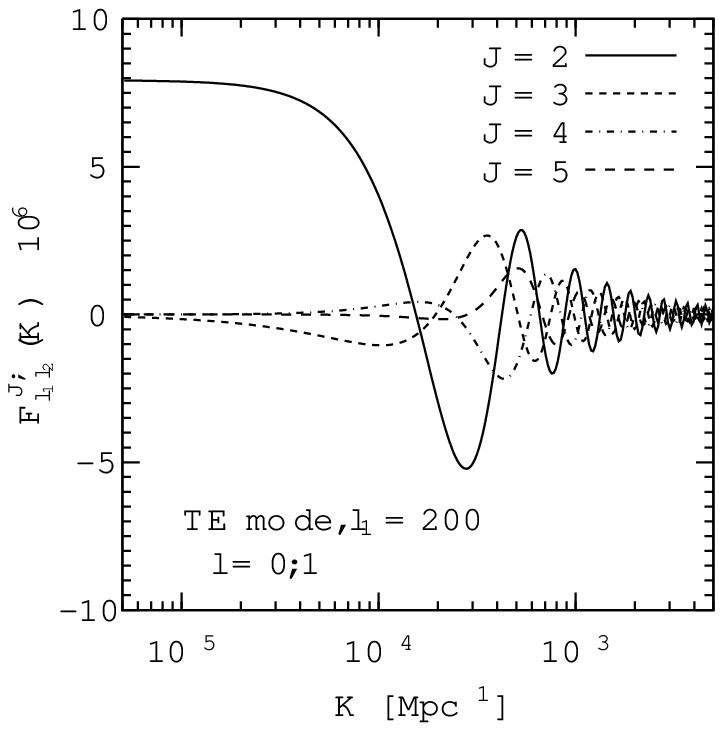}\\
\hspace{-1cm}
\includegraphics[scale=0.9]{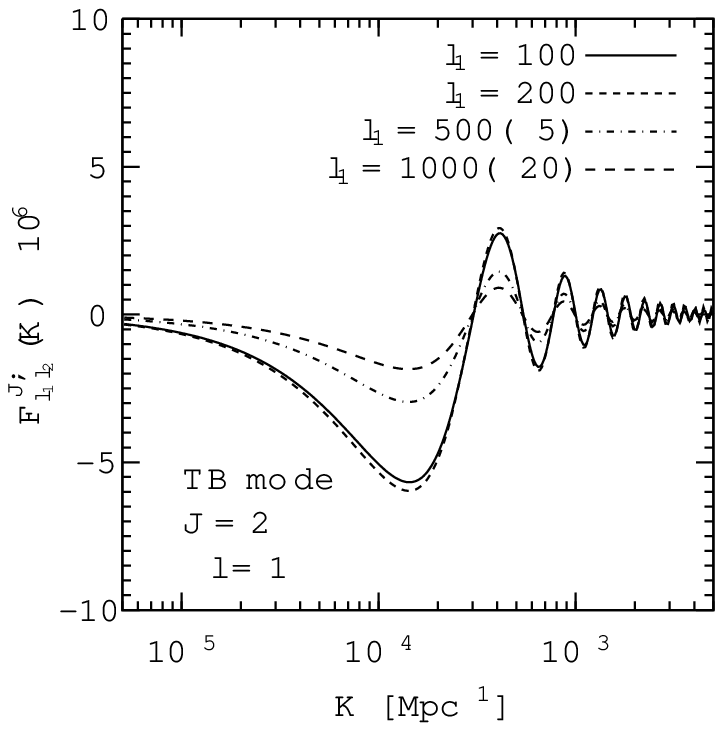}
\hspace{+1cm}
\includegraphics[scale=0.9]{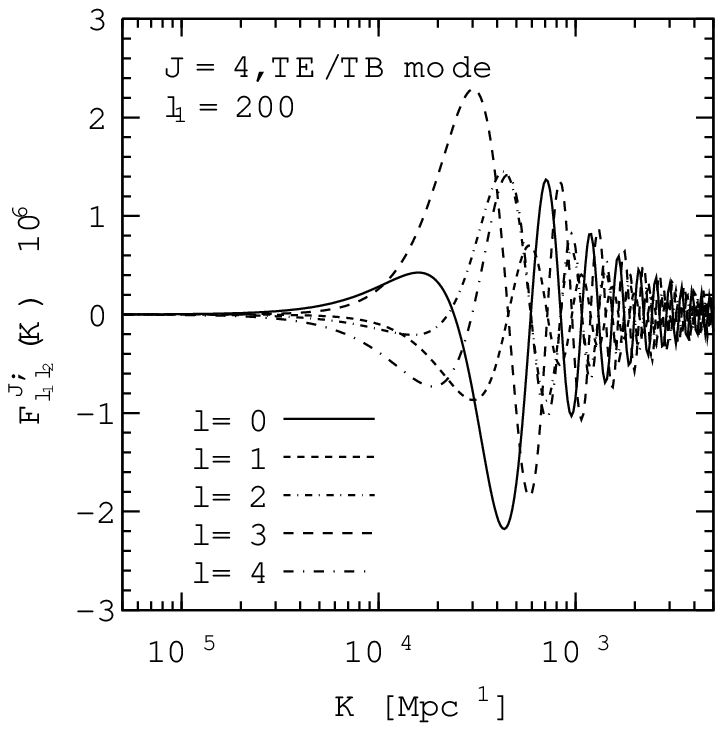}
\caption{Examples of coefficient functions $F^{J,\alpha}_{l_1 l_2}(K)$ with $\Delta l=l_2-l_1$. Note that we plot $-iF^{J,\alpha}_{l_1 l_2}(K)$ for $J=$ odd since $F^{J,\alpha}_{l_1 l_2}(K)$ is imaginary in that case. (a) Top left panel: $F^{J,\alpha}_{l_1 l_2}$ of different polarization types $\alpha=L,VE,VB,TE,TB$ have different peak locations. Parity-even $L,VE,TE$ modes contribute infrared divergence to the quadrupolar $J=2$ BiPoSHs. Parity-odd $VB,TB$ modes are not relevant to this infrared divergence. (b) Top right panel: as $J$ increases, extrema and nodes of $F^{J,\alpha}_{l_1 l_2}$ are systematically shifted to smaller scales, and the amplitudes decrease as well. (c) Bottom left panel: increasing the CMB multipole $l_1$ rescales the amplitude of $F^{J,\alpha}_{l_1 l_2}$, but extrema and nodes are marginally shifted. (d) Bottom right panel: the amplitude varies for different values of $\Delta l$, and the locations of extrema and nodes change slightly.}
\label{fig:Fprofile}
\end{figure*}

We numerically evaluate and tabulate coefficient functions $F^{J,\alpha}_{l_1 l_2} (K)$ from Eq.~(\ref{eq:Fcoef}) for $5\times10^{-6} \mathrm{Mpc}^{-1} \leqslant K \leqslant 5\times 10^{-3} \mathrm{Mpc}^{-1}$, ranging from fossil-field modes that are well outside of the horizon today to modes that were inside of the horizon at the last-scattering surface. Examples of $F^{J,\alpha}_{l_1 l_2} (K)$ are plotted in Fig.~\ref{fig:Fprofile} to show some of the features. In the TAM-wave picture, the oscillatory nature of the $F^{J,\alpha}_{l_1 l_2} (K)$'s can be understood as the value of the TAM wave function on the surface of last scattering. As $K$ increases, spherical waves shrink inward to the origin, and hence nodes and anti-nodes alternate to pass the surface of last scattering. On the other hand, for large scale modes with very small values of $K$, even the first peak is well beyond the last scattering surface, and consequently BiPoSHs vanish in the infrared. Exceptions are the $J=2$ even-parity modes $\alpha=L,VE,TE$, whose wave functions do not vanish at the origin. Therefore, logarithmic infrared divergence arises in even-parity quadrupolar BiPoSHs due to super-horizon TAM modes. In this work, we introduce an infrared cutoff at $K_{\mathrm{min}}=5\times10^{-6} \mathrm{Mpc}^{-1}$, which does not affect the results except for the even-parity $J=2$ BiPoSHs.

\begin{figure}
\centering
\includegraphics[scale=0.9]{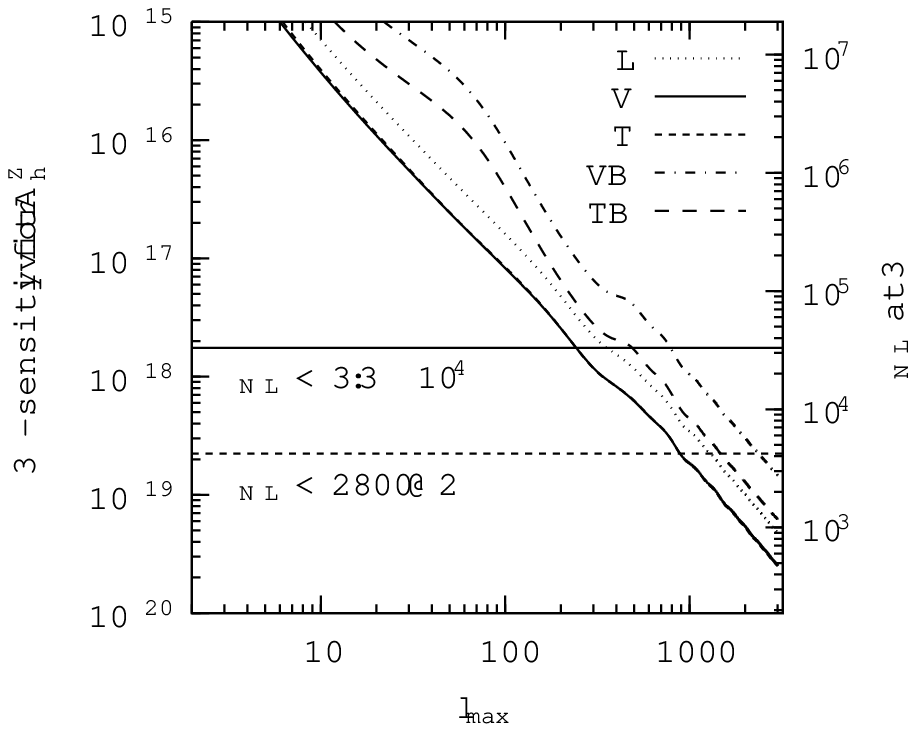}
\caption{Predicted $3\sigma$ sensitivity for the fossil-field reduced amplitude $\mathcal{A}^{Z}_h$ as a function of the maximum multipole $\lmax$. Only BiPoSHs with $J\leqslant 5$ are considered, as they dominate the signal. For the quadrupolar signal $J=2$ from $L,VE,TE$ modes, we cut off the infrared divergence at $K=5\times10^{-6}\mathrm{Mpc}^{-1}$. We relate our results to the primordial trispectrum parameter $\tau_{NL}$ in the local model of non-Gaussianity through $\mathcal{A}^L_h=5.3\times 10^{-23}\tau_{NL}$. We show the {\rm WMAP} 5-year constraint (horizontal solid), and the constraint from the first {\rm Planck} data release (horizontal dashed).}
\label{fig:NoisePlot}
\end{figure}

We calculate sensitivities for $\mathcal{A}^{Z}_h$ at 3$\sigma$ statistical significance (Fig.~\ref{fig:NoisePlot}) as a function of the largest CMB multipole $\lmax$ dominated by the signal. No instrumental noise is assumed for $l\leqslant\lmax$. We consider scalar, vector or tensor fossil fields $Z=L,V,T$ respectively using two strategies. One is to fully exploit the information from both even- and odd-parity BiPoSHs. Since the signal-to-noise is dominated by even-parity BiPoSHs, this optimizes the sensitivity. In reality, however, lensing by gravitational potentials and other late-time mechanisms kick in on small angular scales $l\gtrsim 1000$. They mimic the effect of fossil fields by generating even-parity BiPoSHs as well, making it necessary to disentangle between them. To circumvent the problem, the second strategy is to include only odd-parity BiPoSHs. They provide clean probes of fossils with non-zero spin, with moderate compromise in the overall sensitivity. Even though there is degeneracy between the reduced amplitudes $\mathcal{A}^V_h$ and $\mathcal{A}^T_h$ from measurement of the CMB BiPoSHs, one can still break it with a three-dimensional measurement from galaxy-clustering~\cite{Jeong:2012df}.

The BiPoSH signal is dominated by anisotropic modulation on the largest angular scales, and $J\leqslant 5$ almost saturates the detectability. On the other hand, modulation of small-scale anisotropies (very large $l_1$ and $l_2$) contributes large signal-to-noise to bring down $\sigma^Z_{\mathcal{A}}$ which is found to scale as $\lmax^{-2}$, in agreement with similar studies of CMB trispectra~\cite{Kogo:2006kh, Kamionkowski:2010me}. Such a scaling law is justified analytically in the Sachs-Wolfe limit, valid on large scales. In that case, the radiation transfer functions are approximated by $g^{T}_l(k)=-j_l(kr^*)/3$, where $r^*\approx 14~\mathrm{Gpc}$ is the comoving distance to the last scattering surface. For a given BiPoSH angular scale $J$ and CMB scale $l_1,l_2$ satisfying $L\ll l_1 \approx l_2 \approx l_{12}$ and $L\ll l_3 \approx l_4 \approx l_{34}$, the signal squared can be shown to scale as $l_{12}^{-3}l_{34}^{-3}$, while the noise squared damps out more rapidly as $l_{12}^{-4}l_{34}^{-4}$, so that each estimator $\widehat{\mathcal{A}^{J,Z}_{h,l_1 l_2,l_3 l_4}}$ contributes signal-to-noise squared $\propto l_{12}l_{34}$. Given the constraints $|l_1-l_2|<J$ and $|l_3-l_4|<J$, the number of such estimators scales as $J\sum_{l_{12}\leqslant\lmax}\sum_{l_{34}\leqslant\lmax}$. Therefore, the cumulative signal-to-noise squared is $\mathcal{S}^2\propto J\sum_{l_{12}\leqslant\lmax}\sum_{l_{34}\leqslant\lmax} l_{12} l_{34} \propto \lmax^4$. Nonetheless, at $\lmax\gtrsim 3000$, gravitational lensing and instrumental error take over~\cite{Babich:2004yc}, cutting off the growth in sensitivity.  

To compare our results to the trispectrum parameter $\tau_{NL}$
widely used in the local model of primordial
non-Gaussianity~\cite{Kunz:2001ym,Hu:2001fa,Okamoto:2002ik,Kogo:2006kh,Regan:2010cn},
we consider induced four-point correlations in (b) of
Fig.~\ref{fig:4pt-correlation} by integrating out an
intermediate scalar fossil (dashed line), which mimics an
intrinsic trispectrum. Since the induced trispectrum is
proportional to $\mathcal{A}^{L}_h$, we find the conversion
$\mathcal{A}^L_h=5.3\times 10^{-23}\,\tau_{NL}$.  In this way,
we compare our results to the {\rm WMAP} 5-year bound
$|\tau_{NL}|<3.3\times 10^4$~\cite{Smidt:2010sv} and to the bound $\tau_{NL}<2800$ at $2\sigma$ from the first {\rm Planck} data release~\cite{Ade:2013ydc}. 
Good discovery potentials for {\rm Planck} with $\lmax=2500$ can be achieved, for both
even-/odd-parity BiPoSHs combined and odd-parity BiPoSHs alone. The {\rm Planck} satellite can improve
upon {\rm WMAP}'s sensitivity to $\mathcal{A}^Z_h$ by a factor
$\sim 25$. Planck's sensitivity corresponds to that of a
galaxy-clustering surveys with volume
$V=60\,\mathrm{Gpc}^3\,h^{-3}$ and a resolution
$k_{\mathrm{max}}=0.1\,h\,\mathrm{Mpc}^{-1}$. Thus, Planck's
sensitivity is in principle nearly two orders of magnitude better than that
of the current SDSS-III BOSS survey~\cite{Anderson:2012sa}. Besides, we find good
numerical agreement with the $\tau_{NL}$ forecasts of
Ref.~\cite{Kogo:2006kh}.

\section{Local-departure modulation in real space}
\label{sec:local-modulation-real-space}

There exists, for the local-type scalar-scalar-fossil, a simple and illustrative real-space picture of the effect of the fossil field on primordial perturbations.  To see this, we write,
\be
\label{eq:JK-local-depart}
\Phi(\bfx) = \left( 1 + \frac{\mathcal{B}^{Z}_h}{2 \mathcal{P}_{\Phi}} h^{\alpha}_{ab} (\bfx) \nabla^{-2} \nabla^a \nabla^b \right) \Phi_g(\bfx),\quad \alpha\in Z.
\ee
This reproduces the squeezed limit $K\rightarrow0$ of Eq.~(\ref{eq:JK-fourier}) with the local-departure form Eq.~(\ref{eq:bispectrum-shape}). Here $\Phi_g(\bfx)$ is the gaussian field, and $\mathcal{P}_{\Phi}$ is the dimensionless normalization of the scalar-perturbation power spectrum. Note that the relation between $\Phi(\bfx) $ and $\Phi_g(\bfx)$ is local.

For CMB multipoles $l\lesssim 100$, the temperature is determined primarily by the value of the potential $\Phi$ at the surface of last scattering. Therefore, consider one TAM mode of the fossil field. The distortions to hot and cold spots induced by the fossil field are proportional to the value of the TAM wave function $\Psi^{\alpha,k}_{(JM)ab}(\bfx)$ on the surface of last scattering. This reduces to three-dimensional tensor field that lives on the surface of the two-sphere. This tensor field is a linear combination of the five tensor spherical harmonics $Y^{\beta}_{(JM)}(\bfn),~\beta=L,VE,VB,TE,TB$, as shown in Eq.~(94) of Ref.~\cite{Dai:2012bc} where the comoving radial distance $r$ corresponds to the comoving distance to the surface of last scattering. We clarify here that although these tensor spherical harmonics are only functions of two-dimensional $\bfn$, their tensor value at each $\bfn$ can have components purely perpendicular ($\beta=TE,TB$) or parallel ($\beta=L$) to the line of sight, or even partially perpendicular ($\beta=VE,VB$).

\begin{figure}[ht]
\centering
\includegraphics[scale=0.6]{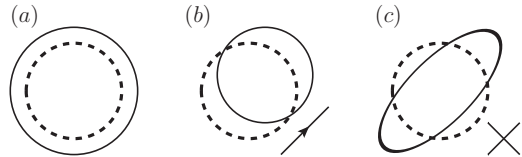}
\caption{Local distortion of hot/cold spots as a result of modulation from tensor harmonics. Intrinsic (dashed) and distorted (solid) isothermal contours are shown. (a) longitudinal harmonic $Y^{L}_{(JM)ab}(\bfn)$; (b) vectorial harmonics $Y^{VE,VB}_{(JM)ab}(\bfn)$; (c) tensorial harmonics $Y^{TE,TB}_{(JM)ab}(\bfn)$.}
\label{fig:local-distort}
\end{figure}

Consider a patch in the sky, which is small relative to the typical variation scale of the fossil field, but large relative to the scale of hot and cold spots. Then according to Eq.~(\ref{eq:JK-local-depart}) the fossil field acts as a constant modulation factor across the patch. From Eq.~(\ref{eq:JK-local-depart}), the departure is proportional to the double gradient of $\Phi_g(\bfx)$, but only the components purely perpendicular to the line of sight can be seen. These then distort isothermal contours on the sky. This implies that local distortion patterns are distinct for each of the five types of tensor harmonics (Fig.~\ref{fig:local-distort}): (1) The longitudinal harmonic $Y^{L}_{(JM)ab}(\bfn)$ is purely radial, so from a projected view onto the sky, isothermal contours expand or shrink in an isotropic way. As a result, only the contrast between hot and cold spots are enhanced or reduced, but the shape and the center remain the same. (2) The vectorial harmonics $Y^{VE,VB}_{(JM)ab}(\bfn)$ establish a preferred direction at given $\bfn$ from a projected view onto the sky. Isothermal contours are shifted when they are perpendicular to that preferred direction, but are unchanged if parallel to it. The outcome is that spots are shifted parallel or anti-parallel to the preferred direction, and hot spots and cold spots shift in the same direction. (3) The purely transverse tensorial harmonics $Y^{TE,TB}_{(JM)ab}(\bfn)$ define two principle directions perpendicular to each other at given $\bfn$ from a projected view onto the sky. Isothermal contours always shift to the hot side along one principle direction, and shift to the cold side along the other principle direction. This leaves the center of the hot and cold spots unchanged, but induces ellipticity by elongating the spot along one of the principle directions and squashing it along the other.

\begin{figure}[ht]
\centering
\includegraphics[scale=0.42]{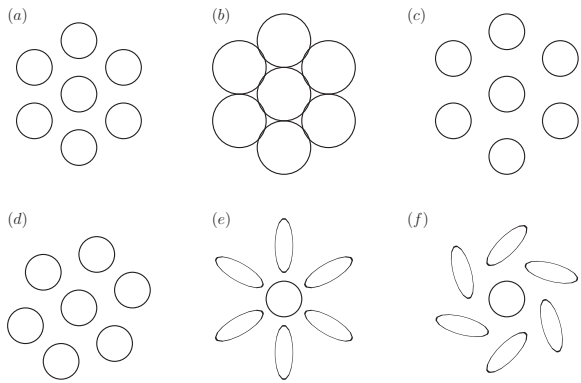}
\caption{Modulation of isothermal contours by different tensor harmonics: (a) intrinsic contours without modulation; (b) longitudinal type; (c) vectorial $E$-type; (d) vectorial $B$-type; (e) tensorial $E$-type; (f) tensorial $B$-type.}
\label{fig:patch-vary}
\end{figure}

On the scale of the fossil-field modulation, the local distortion pattern varies from patch to patch (Fig.~\ref{fig:patch-vary}). Given fossil-field angular-momentum quantum numbers $J$ and $M$, the full sky is divided into regions centered at zeroes of the harmonic. For modulation by the longitudinal harmonic $Y^{L}_{(JM)ab}(\bfn)$, the hot/cold contrast is enhanced in one of the two adjacent regions and is reduced in the other. For modulation by the vectorial $E$-mode harmonic $Y^{VE}_{(JM)ab}(\bfn)$, rings of hot/cold spots around the center of the region either expand away from it, or sink toward it, and an adjacent region has the opposite effect. For the $B$-mode harmonic $Y^{VB}_{(JM)ab}(\bfn)$, rings of hot/cold spots systematically rotate around the center of the region, with opposite rotating directions in adjacent regions. Similarly, for modulation by the tensorial $E$-harmonic $Y^{TE}_{(JM)ab}(\bfn)$, hot/cold spots are elongated along the radial/tangential direction with respect to the center of the region. By contrast, for the $B$-mode harmonic $Y^{TB}_{(JM)ab}(\bfn)$ the direction of elongation is rotated by $45^{\circ}$. Again, in adjacent regions the effect is the opposite.

In the far-field limit, i.e. small fossil-field wavelength relative to the comoving distance to the surface of last scattering, the TAM wavefunction $\Psi^{\alpha,k}_{(JM)ab}(\bfx)$ is asymptotically proportional to the corresponding type of tensor harmonic $Y^{\alpha}_{(JM)ab}(\bfn)$. In that case, longitudinal TAM waves mainly modulate hot/cold contrast, divergence-free vectorial TAM waves mainly induce displacement, and divergence-free tensorial TAM waves mainly generate ellipticity. However, the interesting scales for a bispectrum of the local type are large scales, and a given TAM wavefunction typically projects to different types of tensor harmonics at last scattering and generates different types of real-space distortions simultaneously.

\section{Conclusion}
\label{sec:conclusion}

In this paper, we have investigated the signature of inflation fossils in the anisotropies of the CMB temperature. To take into account the possibilities that the fossil can be a scalar, a vector or a tensor, we have used a generic parameterization for cross-correlations of two scalar perturbation modes as a consequence of the fossil. In harmonic space, non-zero BiPoSH coefficients are induced; in real space, the effects show up as modulations of temperature contrast and displacement and ellipticity of hot and cold spots from patch to patch across the sky. We have provided results for BiPoSH coefficients and BiPoSH power spectra for general inflation fossil scenarios.

As an important example, we have numerically examined the case of a local-type modulation, featuring a scalar-scalar-fossil bispectrum dominated by the squeezed limit and a scale-free fossil power spectrum. By constructing the minimum-variance quadratic estimator, we have found that the combination $\mathcal{A}^Z_h=\mathcal{P}^{Z}_h(\mathcal{B}^{Z}_h)^2$ can be measured. Our numerical results show that modulations on large angular scales dominate the signal, and the sensitivity improves as $\lmax^{-2}$. For an experiment with {\rm Planck}'s sensitivity $\lmax=2500$ and with a combined analysis of both even-parity and odd-parity BiPoSHs, we find detectability at $\mathcal{A}^Z_h / 10^{-20} = 6.6,~3.5,~3.5$ for $Z=L,V,T$ respectively, at 3$\sigma$ significance. On the other hand, odd-parity BiPoSHs alone are clean probes of vector/tensor fossils, with sensitivities $\mathcal{A}^Z_h / 10^{-20} = 20,~8.6$ for $Z=V,T$ respectively. These results are comparable to a sensitivity for the trispectrum parameter in the local model $\tau_{NL}\sim 1250$. They are also significantly better than the sensitivity estimated to be available with current galaxy-clustering surveys.

Even better signal-to-noise is expected if the $E$-mode polarization is also considered (the scalar-scalar-fossil bispectrum we discuss here does not generate or affect the $B$-mode polarization). In that case, the BiPoSHs arising from $TT$-correlation, $EE$-correlation and $TE$-correlation are correlated with each other. Therefore, a correct treatment must involve inverting the covariance matrix. We do not include the full analysis with polarization in this paper, leaving it to future work. Also, it will not be possible to geometrically distinguish between scalar-, vector- and tensor-type fossil field from the CMB. A possible solution might be to scrutinize the shape of the bipolar power spetra. The shape is found to be degenerate between vector and tensor fossils for a primordial bispectrum of local type, but it might be more discriminative for models with primordial bispectra of other shapes. In the latter case, fossil modes smaller than the scale of the last scattering surface can be important, and therefore distince BiPoSH shapes can be induced due to different polarization patterns for the scalar-, vector- and tensor-type fossil TAM wavefunctions in the far-field limit, as discussed in Sec.~\ref{sec:local-modulation-real-space}.

Weak lensing of the CMB by the foreground mass distribution also generates non-zero BiPoSHs and could thus constitute a background for the fossil-field signal. However, the fossil-field signal peaks (at least for a local-type coupling) at $J\lesssim$ few, while the weak-lensing BiPoSHs peak at $J\sim100$ and will have a very precisely-predictable shape.  Vector and tensor fossil fields will moreover be distinguishable by the odd-parity BiPoSHs, which are not induced by lensing by density perturbations. Finally, lensing introduces a CMB-polarization $B$ mode, which can further distinguish it from the effects of a fossil field.

Before closing we comment on the nature of the infrared divergence in the prediction for the amplitude of the quadrupolar ($J=2$) BiPoSHs that arises if the fossil field has a scale-invariant spectrum and an inflaton-inflaton-fossil three-point function of the local type. The predicted amplitude for the $J=2$ BiPoSH then depends logarithmically on the smallest wavenumber $K$ for the fossil field, and thus, on the onset of inflation. A similar divergence arises also in the fossil-field prediction of galaxy clustering~\cite{Jeong:2012df}. The result seems to imply that distance scales well beyond the horizon are having significant effect on observables within the horizon. A similar logarithmic divergence has been discussed in
Refs.~\cite{Giddings:2010nc,Giddings:2011zd} where it was argued
that the tensor-scalar-scalar three-point correlation in
slow-roll inflation may give rise to an power
quadrupole that could be probed observationally
\cite{Pullen:2007tu,Pullen:2010zy}. One the other hand, in Refs.~\cite{Gerstenlauer:2011ti,Tanaka:2011aj} infrared-safe correlation functions are constructed and are argued to be representative of the statistics in the local Hubble patch. We believe it is important to fully understand this effect, as it occurs from the scalar-scalar-tensor three-point function that arises in single-field slow-roll inflation~\cite{Maldacena:2002vr}, even without the introduction of additional new field.

\begin{acknowledgments}
We thank Fabian Schmidt for useful discussions.
This work was supported by DoE SC-0008108 and NASA NNX12AE86G.
\end{acknowledgments}


\begin{thebibliography}{99}
\bibliographystyle{unsrt}

\bibitem{InflationModelReview}
  D.~H.~Lyth and A.~Riotto,
  Phys.\ Rept.\  {\bf 314}, 1 (1999)
  [hep-ph/9807278].

\bibitem{DBI-Inflation}
  G.~R.~Dvali and S.~H.~H.~Tye,
  Phys.\ Lett.\ B {\bf 450}, 72 (1999)
  [hep-ph/9812483];\
  S.~Kachru, R.~Kallosh, A.~D.~Linde, J.~M.~Maldacena, L.~P.~McAllister and S.~P.~Trivedi,
  JCAP {\bf 0310}, 013 (2003)
  [hep-th/0308055].\

\bibitem{InflationExtraScalar}
  D.~Baumann and D.~Green,
  Phys.\ Rev.\ D {\bf 85}, 103520 (2012)
  [arXiv:1109.0292 [hep-th]].
    A.\ D.\ Linde,
    Phys.\ Lett.\ {\bf B 259}, 38 (1991);\
    A.\ Linde,
    Phys.\ Rev.\ {\bf D 49} 748, (1994)
    [astro-ph/9307002];\
    E.\ J.\ Copeland, A.\ R.\ Liddle, D.\ H.\ Lyth, E.\ D.\ Stewart and D.\ Wands,
    Phys.\ Rev.\ {\bf D 49},\ 6410\ (1994)
    [astro-ph/9401011];\
    A.\ R.\ Liddle, A.\ Mazumdar and F.\ E.\ Schunck,
    Phys.\ Rev.\ {\bf D 58},\ 061301\ (1998)
    [astro-ph/9804177].\

\bibitem{InflationExtraVector}
    K.\ Dimopoulos, M.\ Karciauskas, D.\ H.\ Lyth and Y.\ Rodriguez,
    JCAP\ {\bf 0905}, 013 (2009)
    [arXiv:0809.1055];\
    A.\ Golovnev and V.\ Vanchurin,
    Phys.\ Rev.\ {\bf D 79},\ 103524 (2009);\
    N.\ Bartolo, E.\ Dimastrogiovanni, S.\ Matarrese, and A.\ Riotto,
    J.\ Cosmol.\ Astropart.\ Phys.\ {\bf 11}, 028 (2009).\

\bibitem{Jimenez:2009ai} 
  J.~Beltran Jimenez and A.~L.~Maroto,
  Phys.\ Rev.\ D {\bf 80}, 063512 (2009)
  [arXiv:0905.1245].

\bibitem{InflationGW}
    V.\ A.\ Rubakov, M.\ V.\ Sazhin and A.\ V.\ Veryaskin,
    Phys.\ Lett.\ {\bf B 115}, 189\ (1982);
    R.\ Fabbri and M.\ D.\ Pollock,
    Phys.\ Lett.\ {\bf B 125}, 445 (1983);
    L.\ F.\ Abbott and M.\ B.\ Wise,
    Nucl.\ Phys.\ {\bf B 244}, 541 (1984);
    B.\ Allen,
    Phys.\ Rev.\ {\bf D 37}, 2078 (1988);
    L.\ P.\ Grishchuk,
    Phys.\ Rev.\ Lett.\ {\bf 70}, 2371 (1993).

\bibitem{Maldacena:2002vr} 
  J.~M.~Maldacena,
  JHEP {\bf 0305}, 013 (2003)
  [astro-ph/0210603].

\bibitem{Seery:2008ax} 
  D.~Seery, M.~S.~Sloth and F.~Vernizzi,
  JCAP {\bf 0903}, 018 (2009)
  [arXiv:0811.3934].

\bibitem{Giddings:2010nc} 
  S.~B.~Giddings and M.~S.~Sloth,
  JCAP {\bf 1101}, 023 (2011)
  [arXiv:1005.1056 [hep-th]].

\bibitem{Giddings:2011zd} 
  S.~B.~Giddings and M.~S.~Sloth,
  Phys.\ Rev.\ D {\bf 84}, 063528 (2011)
  [arXiv:1104.0002].

\bibitem{Komatsu:2001rj} 
  E.~Komatsu and D.~N.~Spergel,
  Phys.\ Rev.\ D {\bf 63}, 063002 (2001)
  [astro-ph/0005036].

\bibitem{Babich:2004yc} 
  D.~Babich and M.~Zaldarriaga,
  Phys.\ Rev.\ D {\bf 70}, 083005 (2004)
  [astro-ph/0408455].

\bibitem{Bartolo:2004if} 
  N.~Bartolo, E.~Komatsu, S.~Matarrese and A.~Riotto,
  Phys.\ Rept.\  {\bf 402}, 103 (2004)
  [astro-ph/0406398].

\bibitem{Kogo:2006kh} 
  N.~Kogo and E.~Komatsu,
  Phys.\ Rev.\ D {\bf 73}, 083007 (2006)
  [astro-ph/0602099].

\bibitem{Smidt:2010ra} 
  J.~Smidt, A.~Amblard, C.~T.~Byrnes, A.~Cooray, A.~Heavens and D.~Munshi,
  Phys.\ Rev.\ D {\bf 81}, 123007 (2010)
  [arXiv:1004.1409].

\bibitem{Jeong:2012df} 
  D.~Jeong and M.~Kamionkowski,
  Phys.\ Rev.\ Lett.\  {\bf 108}, 251301 (2012)
  [arXiv:1203.0302].

\bibitem{Bardeen:1980kt} 
  J.~M.~Bardeen,
  Phys.\ Rev.\ D {\bf 22}, 1882 (1980).

\bibitem{Dai:2012bc} 
  L.~Dai, M.~Kamionkowski and D.~Jeong,
  Phys.\ Rev.\ D {\bf 86}, 125013 (2012)
  [arXiv:1209.0761].

\bibitem{Dai:2012ma} 
  L.~Dai, D.~Jeong and M.~Kamionkowski,
  Phys.\ Rev.\ D in press,
  [arXiv:1211.6110].

\bibitem{Hajian:2004zn} 
  A.~Hajian, T.~Souradeep and N.~J.~Cornish,
  Astrophys.\ J.\  {\bf 618}, L63 (2004)
  [astro-ph/0406354].

\bibitem{Book:2011na} 
  L.~G.~Book, M.~Kamionkowski and T.~Souradeep,
  Phys.\ Rev.\ D {\bf 85}, 023010 (2012)
  [arXiv:1109.2910].

\bibitem{Komatsu:2010fb} 
  E.~Komatsu {\it et al.}  [WMAP Collaboration],
  Astrophys.\ J.\ Suppl.\  {\bf 192}, 18 (2011)
  [arXiv:1001.4538].

\bibitem{Lewis:1999bs} 
  A.~Lewis, A.~Challinor and A.~Lasenby,
  Astrophys.\ J.\  {\bf 538}, 473 (2000)
  [astro-ph/9911177].

\bibitem{Kamionkowski:2010me} 
  M.~Kamionkowski, T.~L.~Smith and A.~Heavens,
  Phys.\ Rev.\ D {\bf 83}, 023007 (2011)
  [arXiv:1010.0251].

\bibitem{Kunz:2001ym} 
  M.~Kunz, A.~J.~Banday, P.~G.~Castro, P.~G.~Ferreira and K.~M.~Gorski,
  [astro-ph/0111250].

\bibitem{Hu:2001fa} 
  W.~Hu,
  Phys.\ Rev.\ D {\bf 64}, 083005 (2001)
  [astro-ph/0105117].

\bibitem{Okamoto:2002ik} 
  T.~Okamoto and W.~Hu,
  Phys.\ Rev.\ D {\bf 66}, 063008 (2002)
  [astro-ph/0206155].

\bibitem{Regan:2010cn} 
  D.~M.~Regan, E.~P.~S.~Shellard and J.~R.~Fergusson,
  Phys.\ Rev.\ D {\bf 82}, 023520 (2010)
  [arXiv:1004.2915].

\bibitem{Smidt:2010sv} 
  J.~Smidt, A.~Amblard, A.~Cooray, A.~Heavens, D.~Munshi and P.~Serra,
  [arXiv:1001.5026].

\bibitem{Ade:2013ydc} 
  P.~A.~R.~Ade {\it et al.}  [Planck Collaboration],
  arXiv:1303.5084 [astro-ph.CO].

\bibitem{Anderson:2012sa} 
  L.~Anderson, E.~Aubourg, S.~Bailey, D.~Bizyaev, M.~Blanton, A.~S.~Bolton, J.~Brinkmann and J.~R.~Brownstein {\it et al.},
  Mon.\ Not.\ Roy.\ Astron.\ Soc.\  {\bf 428}, 1036 (2013)
  [arXiv:1203.6594].

\bibitem{Pullen:2007tu} 
  A.~R.~Pullen and M.~Kamionkowski,
  Phys.\ Rev.\ D {\bf 76}, 103529 (2007)
  [arXiv:0709.1144 [astro-ph]].

\bibitem{Pullen:2010zy} 
  A.~R.~Pullen and C.~M.~Hirata,
  JCAP {\bf 1005}, 027 (2010)
  [arXiv:1003.0673 [astro-ph.CO]].

\bibitem{Gerstenlauer:2011ti} 
  M.~Gerstenlauer, A.~Hebecker and G.~Tasinato,
  JCAP {\bf 1106}, 021 (2011)
  [arXiv:1102.0560 [astro-ph.CO]].

\bibitem{Tanaka:2011aj} 
  T.~Tanaka and Y.~Urakawa,
  JCAP {\bf 1105}, 014 (2011)
  [arXiv:1103.1251].


\end{thebibliography}

\end{document}